\documentclass{kapproc} 
\usepackage{amssymb,amsfonts,color,colordvi,epsf,latexsym}
\usepackage[dvips]{graphics} 

 \kluwerbib          

\def\complex       {{\mathbb C}}
\def\calc          {{\cal C}}
\def\calh          {{\cal H}}
\def\chii          {\raisebox{.15em}{$\chi$}}  
\renewcommand\cite[2]{[#1]}
\def\eq            {\,{=}\,}
\def\Hom           {{\rm Hom}}
\def\id            {\mbox{\sl id}}
\def\iN            {\,{\in}\,}    
\def\oti           {\,{\otimes}\,}
\def\Oti           {{\otimes}}
\def\reals         {{\mathbb R}}
\def\tft           {topological field theory}
\def\tfts          {topological field theories}
\def\zet           {{\mathbb Z}}

\begin{document}
\thispagestyle{empty}
\begin{flushright} \begin{tabular}{rl} {~} & {~} \\[-12.3mm]
{\sf hep-th/0110158}\\[1mm]{\sf PAR-LPTHE 01-51}\\[1mm]
{\sf October 2001}\\[8mm]{} \end{tabular} \end{flushright}


\begin{center} 
{\Large\bf CONFORMAL BOUNDARY CONDITIONS}\\[3mm]
{\Large\bf AND 3D TOPOLOGICAL FIELD THEORY} \\[1mm]
\end{center}
\begin{minipage}{4cm}
\author{J\"urgen Fuchs}
\affil{Institutionen f\"or fysik~~~~{}\\
Universitetsgatan 1\\ S\,--\,651\,88\, Karlstad}
\end{minipage}
\begin{minipage}{8cm}
\author{Ingo Runkel \ and \ Christoph Schweigert}
\affil{LPTHE, Universit\'e Paris VI~~~{}\\
4 place Jussieu\\ F\,--\,75\,252\, Paris\, Cedex 05}
\end{minipage}

\begin{abstract}
Topological field theory in three dimensions provides a powerful tool
to construct correlation functions and to describe boundary conditions in
two-dimensional conformal field theories. 
\end{abstract}

\begin{keywords}
Conformal field theory, topological field theory, boundary conditions
\end{keywords}


\section{Introduction}

There are numerous applications of two-dimensional conformal field theories 
on manifolds with boundaries. They range from impurities in systems of condensed 
matter physics to D-branes in string theory. The present contribution explains 
an approach to correlators in such theories that is based on a special 
instance of a ``holographic correspondence": The spaces of conformal blocks can 
be understood both as spaces of physical states of a three-dimensional topological
field theory (TFT) and as spaces of (pre-)correlators of a two-dimensional
conformal field theory (CFT).

Cardy's formula
  $$  |a\rangle\!\rangle = \sum_i \, \frac{S_{ai}}{S_{0i}} \,\, |i\rangle 
  \label{cardy} $$ 
expressing a boundary state $|a\rangle\!\rangle$ in terms of Ishibashi states 
$|i\rangle$ describes a symmetry preserving boundary condition $a$ for a
CFT with charge conjugation modular invariant. The appearance of the coefficient
$S_{ai}/S_{0i}$ in Cardy's expression is a {\em model-independent\/} 
feature. It involves the modular matrix $S$, which also appears in the 
corresponding TFT as the invariant of the Hopf link in $S^3$.
Roughly speaking, our results provide such topological information
for correlators on orientable world sheets of arbitrary topology, possibly
with boundary, for (rational) CFTs with general modular invariant. At the same
time, we obtain a simple description of conformally invariant boundary
conditions.

This contribution is organized as follows. In Section 2 we 
introduce a double cover of the world sheet to formulate
the problem of finding correlation functions. Section 3 gives a short 
introduction to TFTs in three dimensions and the description of
Moore-Seiberg data in terms of tensor categories. Section 4 presents
our construction of the correlators, while Section 5 explains the
consistency of our ansatz and lists some explicit results. Section 6
contains the conclusions.  

\section{CFT correlators versus conformal blocks}

Let us formulate concisely the issue we want to address:
\\[-1.14em]{}\begin{center} \fbox{\begin{minipage}{11.7cm}
A correlation function on a world sheet $X$ is fully determined by the choice of 
a vector in the space of conformal blocks associated to a double cover $\hat X$
of $X$. \\
These vectors must obey factorization constraints and 
give modular invariant correlation functions.
\end{minipage}} \end{center}

To explain this idea in more detail, we start with its geometrical aspects:
The doubling of the world sheet is an incarnation of the standard concept
of mirror charges -- to construct correlators on, say,
the disc, we consider conformal blocks on the sphere $\complex{\rm P}^1$. 
The original world sheet $X$ is obtained from the double
cover $\hat X$ as the quotient 
  $$ X = \hat X /\sigma $$
by an anticonformal involution $\sigma$.
The fixed points of $\sigma$ correspond to the boundary points of $X$. 
When $X$ is orientable and closed, $\hat X$ consists of two 
disjoint copies of $X$, endowed with opposite orientation.

The double cover $\hat X$ is oriented and closed and has thus enough structure 
to support a {\em chiral\/}
CFT. A chiral CFT associates to a collection $\{\calh_q\}_{q\in I}$ of
inequivalent irreducible representations of the chiral algebra, labelled by 
some index set $I$ (which is finite when the theory is rational), and to 
two-manifolds with arcs carrying labels from $I$ vector spaces of conformal 
blocks. More precisely, given an closed, oriented two-manifold $\hat X$ 
with insertion points $\vec p\eq(p_1,...\,,p_m)$ and
labels $\vec\lambda\eq(\lambda_1,...\,,\lambda_m)$ in $I$, the vector space of
conformal blocks on $\hat X$ consists of linear functionals
  $$ \beta_{\vec\lambda,\hat X}: \quad
  \calh_{\lambda_1}\,{\otimes}\cdots{\otimes}\,\calh_{\lambda_m} \to\complex
  \, . $$
They should be thought of as {\em pre-}correlators which yield the conformal 
blocks of primary fields as well as descendants by evaluation at the 
corresponding state vector:
  $$  \langle \Phi_{\lambda_1}(v_1,p_1) \,{\cdots}\, \Phi_{\lambda_m}(v_m,p_m)
  \rangle^{}_{\!\hat X} =
  \beta_{\vec\lambda,\hat X}(v_1\,\Oti\cdots\Oti\, v_m) \, . $$ 
A special case is provided by two-point blocks on the sphere, with
insertions at $z\eq0$ and $z\eq\infty$. In this case the functional
  $$ \beta:\quad \calh_\lambda \oti\calh_{\lambda^{\!\vee}_{}} \to \complex $$
is frequently described by a generalized coherent closed string state
$|\lambda\rangle\iN\calh_\lambda\oti\calh_{\lambda^\vee}$ (which in the 
context of boundary conditions is also known as Ishibashi state) such that
  $$ \beta(v_1\Oti v_2) = \langle \lambda \,|\, v_1\Oti v_2\rangle \, . $$

The correlation function on $X$ should be regarded as a vector in the space of
pre-correlators on $\hat X$. If $X$ is closed and orientable, this
amounts to the well-known statement that the correlation functions are
bilinear combinations of left movers and right movers. The present formulation
of the problem has the advantage that it works on all world sheets $X$.

\section{Topological field theory}

The double cover $\hat X$ accomplishes a geometric separation of
left and right movers and therefore gives rise to a {\em chiral\/} theory
on $\hat X$. It is a fruitful general idea that a chiral system
can be understood as the edge system of a higher dimensional
theory. This is already implicit in the usual treatment of chiral
anomalies which uses forms of degree higher than the dimension of the relevant
space-time. The most direct analogue, however, is provided by the quantum Hall
effect: In the scaling limit, the 2+1-dimensional bulk is described
by a TFT, while a chiral CFT on the boundary describes the edge system,
including the edge currents.

Motivated by this analogy, we wish to regard the double $\hat X$ as the boundary
of an appropriate 3-manifold $M_X$ and find a TFT associated to the chiral
CFT. For the latter question, a complete machinery is available: Given
Moore-Seiberg data of the chiral theory (such as braiding and fusing matrices 
and fractional parts of conformal weights) the construction of \cite{13}{TUra} 
provides us with a TFT.

We describe Moore-Seiberg data in terms of modular tensor categories.
This language, albeit not absolutely standard, has two crucial advantages. 
First, it leads to a powerful graphical calculus in terms of ribbon graphs that
correspond to framed Wilson lines. The simplest example of a modular
tensor category is the category of finite-dimensional vector spaces; indeed, 
it is fair to say that modular tensor categories are a natural generalization 
of the category of vector spaces. The second advantage of a systematic use of 
modular tensor categories is that we can continue to use standard
algebraic and representation theoretic tools and intuition to deal with
Moore-Seiberg data.

For the purposes of this summary it is sufficient to think of a modular
tensor category as the category $\calc$ of representations of a chiral algebra.
Fusion provides a notion of a tensor product $\otimes$ on $\calc$, while
conjugate representations give rise to a so-called duality. Moreover,
the statistics of field theories in two dimensions is encoded by a braiding, 
i.e.\ an isomorphism $c_{V,W}$ that interchanges two representations in a 
tensor product, $c_{V,W}\iN\Hom(V\Oti W,W\Oti V)$, but does not necessarily square
to one. Finally, the exponentionated conformal weight gives rise to a twist.

{}From these data one can build a TFT. The basic feature of three-dimensional
TFT is that it provides a {\em modular functor\/}: To geometric data it
associates algebraic structures. Concretely, it associates vector spaces --
the spaces $\calh(\hat X)$ of conformal blocks -- to 
two-dimensional manifolds $\hat X$, and to three-manifolds, endowed with
somewhat more structure, it assigns linear maps between such vector spaces.

More precisely, conformal blocks are associated to {\em extended
surfaces\/} -- two-dimen\-si\-o\-nal closed oriented manifolds with a finite
collection of small arcs. Each arc carries a label from a set $I$.
In our application, these are primary fields, or equivalently, irreducible
representations of a chiral algebra. Moreover, we must
choose a Lagrangian subspace of $H_1(\hat X,\reals)$. We will 
suppress these auxiliary data in our discussion.

The linear maps are associated to cobordisms $(M,\partial_- M,
\partial_+ M)$. Here $M$ is a three-manifold whose boundary $\partial M$
has been decomposed in two disjoint subsets $\partial_\pm M$, each of
which can be empty. Moreover, a ribbon graph has to be chosen in $M$.
After choosing Lagrangian subspaces in $H_1(\partial_\pm M,\reals)$,
the two spaces $\partial_\pm M$ become extended surfaces. The 
linear map associated to the cobordism is then
  $$ Z(M,\partial_- M,\partial_+ M) : \quad
  \calh(\partial_- M) \to \calh(\partial_+ M) \,. $$
In the application of our interest, we always take $\partial_- M$ to be empty. 
Using the fact that $\calh(\emptyset)\,{=}\,\complex\,$, we then obtain a map
  $$ Z(M,\partial M): \quad \complex \to \calh(\partial M) \,, $$
in other words, a line in the vector space $\calh(\partial M)$ of conformal 
blocks. The image $Z(M,\partial M)1$ of the number 1 under this map then 
specifies a vector in $\calh(\partial M)$.

\section{Correlators} 

Topological field theory thus provides a manageable way to describe explicitly
elements in the spaces of conformal blocks, a task that is very difficult
in other approaches to these spaces. As explained in Section 2, this is
precisely the tool we need in order to determine correlation functions. The idea
is to find a three-manifold $M_X$ with boundary $\hat X$ and a Wilson graph 
in $M_X$ such that
  $$ Z(M_X,\emptyset,\hat X): \quad \complex \to\calh(\hat X) $$
gives us the correlation function.

We start with the geometric part of the prescription. The connecting
three-manifold \cite{11,3,4}{hora5,fffs2,fffs3} $M_X$ is defined as the quotient
of the product of the double $\hat X$ with the interval $[-1,1]$, modulo the 
combined action of $\sigma$ on $\hat X$ and $t\,{\mapsto}\,{-}t$ on the interval:
  $$ M_X = \mbox{\Large(} \hat X \times [-1,1] \mbox{\Large)} /\, \zet_2 \,. $$
If $X$ is closed and orientable, $M_X$ is a cylinder over $X$.
For $X$ the disc, the double $\hat X$ is the sphere, and $M_X$ is the
full ball bounded by $\hat X$. 

The world sheet $X$ has a natural embedding $\iota$ into $M_X$: A point 
$p$ in the interior of $X$ with preimages $p^\pm$ on $\hat X$ is mapped to 
$(p^+,0)\,{\sim}\,(p^-,0)\iN$\linebreak[0]$M_X$. Every component of the boundary 
of $X$ gives rise to a circular line of fixed points of the $\zet_2$-action
in $M_X$ and is mapped by
$\iota$ to this line. This geometric construction can be summarized as follows:
We have doubled the the world sheet $X$ to obtain $\hat X$ and fattened
it to a connecting three-manifold $M_X$ with boundary $\hat X$.

\medskip

The next step -- to endow $M_X$ with a Wilson graph -- is much more subtle.
In particular, at this point information on the modular invariant partition 
function of the conformal field theory has to enter. Indeed, for a given chiral
CFT, and hence a given tensor category $\calc$, different partition
functions can exist, and they give rise to different CFTs.

The torus partition function can be written as a bilinear combination 
  $$ Z(\tau) = \sum_{i,j\in I} Z_{ij}\, \chii_i(\tau)\, \chii_j(\tau)^* $$
of the characters
with non-negative integral coefficients $Z_{ij}$. Modular invariance
constitutes a strong constraint on the field content of CFTs (in theories of 
closed strings it even implies the absence of anomalies), and as a consequence 
much work has been spent to classify modular invariants. It comes therefore
as a deception that this classification problem does not have a direct 
physical meaning, as unphysical modular invariants exist. As a consequence, we 
should not, and will not, use the specification of a modular invariant 
partition function as the additional input. Instead \cite9{fuRs}, we specify a 
symmetric special Frobenius algebra in the tensor category $\calc$ that 
formalizes the Moore-Seiberg data. This implicitly also determines a modular 
invariant.

To explain this point, we consider modular invariants of extension type.
In this case the vacuum of the extended theory is a particular {\em reducible}
sector of the original theory. It corresponds to an object $A$
in $\calc$, but this objects inherits additional structure from the
vacuum of the extended theory: The associative OPE induces an associative
product on $A$, which is a morphism $m\iN\Hom(A\oti A,A)$. 
There is also a co-algebra structure, in particular a coproduct
$\Delta\iN\Hom(A,A\oti A)$. Product and co-product are subject to a
number of axioms. Here we highlight just two of them (for the full list see 
\cite9{fuRs}). First, the crossing symmetry between s-channel and t-channel
is taken into account by demanding $A$ to be a {\em Frobenius algebra\/}, 
i.e.\ to satisfy
  $$ (\id_A\oti m) \circ (\Delta\oti\id_A) =
  \Delta\circ m = (m\oti\id_A) \circ (\id_A\oti\Delta) \,.  $$
Pictorially, this is represented as follows:
  $$  \begin{picture}(160,53)(0,0) \put(0,0)   {\begin{picture}(0,0)(0,0)
              \scalebox{.29}{\includegraphics{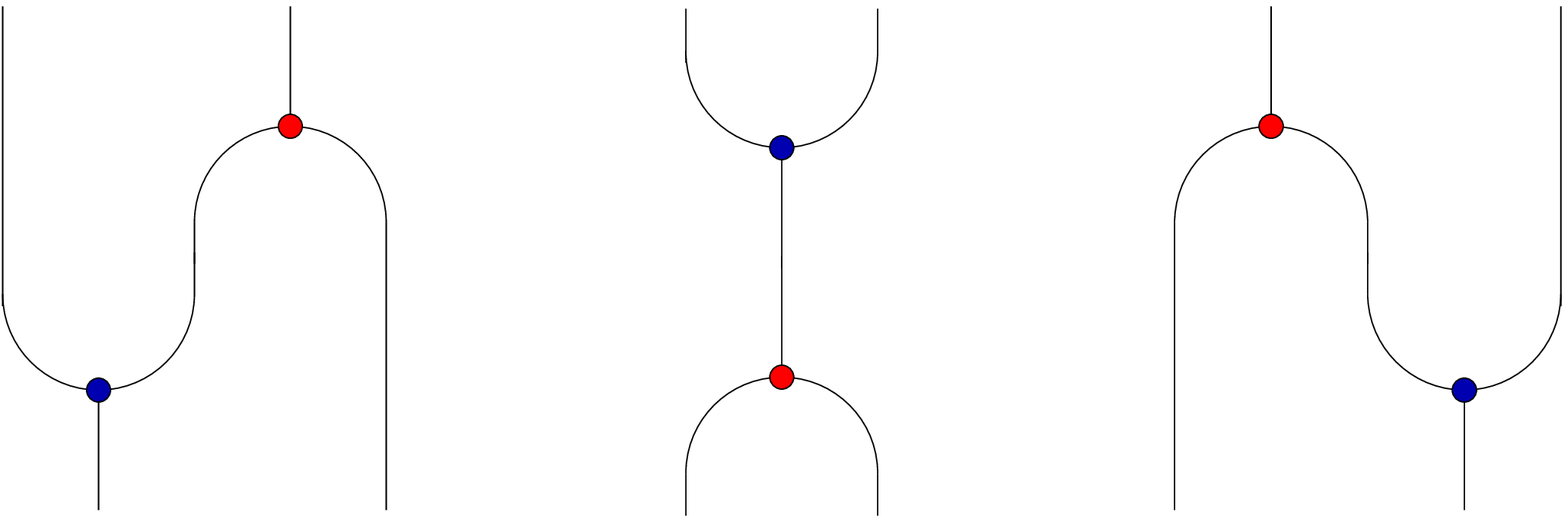}} \end{picture}}
  \put(57,25) {$=$} \put(105,25){$=$} \end{picture} $$
Second, the property that $A$ is special includes the relation
  $$ m\circ \Delta = \id_A  \, . $$
Let us give a few examples for symmetric special Frobenius algebras. 
The relevant algebra for a simple current modular invariant based on
a group $\calh$ of simple currents is
  $$ A = \bigoplus_{J\in\calh} J \,.$$
This object admits inequivalent algebra structures which differ by
elements in $H^2(\calh,\complex^\times)$, corresponding to different choices of
discrete torsion. We stress, however,
that that our formalism treats simple current 
modular invariants and exceptional modular invariants on the same footing.
For instance, the exceptional invariants of the sl$(2)$ WZW theory of type 
$E_6$ (at level 10), $E_7$ (level 16) and $E_8$ (level 28) are obtained 
from the algebras $A\eq0\,{\oplus}\,6$, $A\eq0\,{\oplus}\,8\,{\oplus}\,16$, and 
$A\eq0\,{\oplus}\,10\,{\oplus}\,18\,{\oplus}\,28$, respectively \cite{12}{kios}.

\medskip

The next step in the construction of correlators is to study representations 
of $A$. It turns out that $A$-representations have 
a direct physical interpretation: they correspond to boundary conditions.
A representation theory analogous to the one for vector spaces exists also
for algebra objects in the tensor categories that describe Moore-Seiberg data. 
In particular, there is the notion of an irreducible representation; 
it corresponds to an elementary boundary condition. For rational
CFTs, reducible representations are always fully reducible; they correspond to 
boundary conditions with non-trivial Chan-Paton multiplicities. Standard tools
from representation theory like induced representations or reciprocity
theorems, which generalize to tensor categories (see e.g.\ 
\cite{12,7}{kios,fuSc16}), can be used to work out the list of boundary
conditions in many concrete examples and serve as a rigorous justification of
the procedure proposed in \cite6{fuSc14}. 

\smallskip

We can now finally give a prescription for the Wilson graph in the connecting
manifold $M_X$. It beautifully combines the construction of 
\cite{3,4}{fffs2,fffs3} with structures familiar \cite{10}{fuhk} from
lattice TFTs in {\em two\/} dimensions: \\[.2em]
(1) Each bulk insertion point $p_\ell$ on $X$ has two preimages $\hat p_\ell
    ^\pm$ on $\hat X$, which are joined by an interval in 
    $M_X$. We put a Wilson line along each such interval. The image of
    $X$ in $M_X$ intersects this interval in a unique point,
    which we identify with $p_\ell\in X$. \\[.1em]
(2) Each component of the boundary of $X$ gives rise to a line of fixed points
    of $M_X$ under the $\zet_2$-action. We place a circular Wilson line along
    each such line of fixed points. \\[.1em]
(3) Boundary insertion points $q_\ell$ on $X$ have a unique preimage
    $\hat q_\ell$ on $\hat X$. We join $\hat q_\ell$ by a short Wilson line to
    the image of $q_\ell$ in $M_X$, which results in a trivalent vertex on the
    relevant circular Wilson line. \\[.1em]
(4) We place a Wilson line along every edge of an (arbitrarily chosen)
    triangulation of the image of $X$ in $M_X$. For each bulk insertion we
    join the perpendicular Wilson line in $p_\ell$ by an additional line to 
    an interior point of an arbitrary edge of the face to which $p_\ell$ 
    belongs.  Similarly one joins the triangulation to every segment of the 
    circular boundary lines.
\\[.3em]
This is the geometric part of the prescription. In addition we must
decorate all Wilson lines with labels specifying the corresponding
object of the category and choose couplings (morphisms) for the 
vertices. Our construction is inspired by the situation in two-dimensional 
lattice TFTs and contains that case as a special example. Here we merely 
sketch the idea; for more details see \cite9{fuRs}: \\[.2em]
(5) The triangulation and the short Wilson lines that connect the bulk 
    insertion points $p_\ell$ to the triangulation are labelled by the object
    $A$. \\[.1em]
(6) To vertices that join three $A$-lines we assign morphisms that are
    constructed from the product $m$ on $A$ and from an isomorphism between $A$
    and its dual $A^{\!\vee}$. \\[.1em]
(7) To characterize the bulk field inserted at $p_\ell$, we must specify two
    irreducible objects $j_\ell^\pm$; they label the Wilson lines that start
    from the points $\hat p_\ell^\pm$ on $\hat X$. To account for the coupling
    to the short $A$-lines, we need as a third datum for a bulk field a
    morphism in $\Hom(A\oti j_\ell^+, j_\ell^{-\vee})$. 
(Physical bulk fields correspond only to a subspace of these couplings.
Couplings in a complement of this subspace completely decouple in all amplitudes.)
 \\[.1em]
(8) Each segment of the circular Wilson line corresponds to a boundary condition;
    it is to be labelled by an $A$-representation $M$. Wilson lines of the
    triangulation that end on such a boundary segment result in a trivalent
    vertex, to which we assign the representation morphism for $M$.
 \\[.1em]
(9) The boundary fields have a single chiral label $k_\ell$, which labels the 
    short Wilson line from $\hat q_\ell$ to $q_\ell$. The trivalent vertex 
    that is formed by this Wilson line and the two adjacent boundary 
    conditions $M,N$ requires the
    choice of a coupling in (a subspace of) $\Hom(M\Oti k_\ell,N)$. \\[.3em]
When $X$ is the disk with three bulk 
\\and three boundary insertions, the\\ picture then looks as follows:
  $$   \begin{picture}(0,110)(-8,0)
  \put(0,0)   {\begin{picture}(0,0)(0,0)
              \scalebox{.38}{\includegraphics{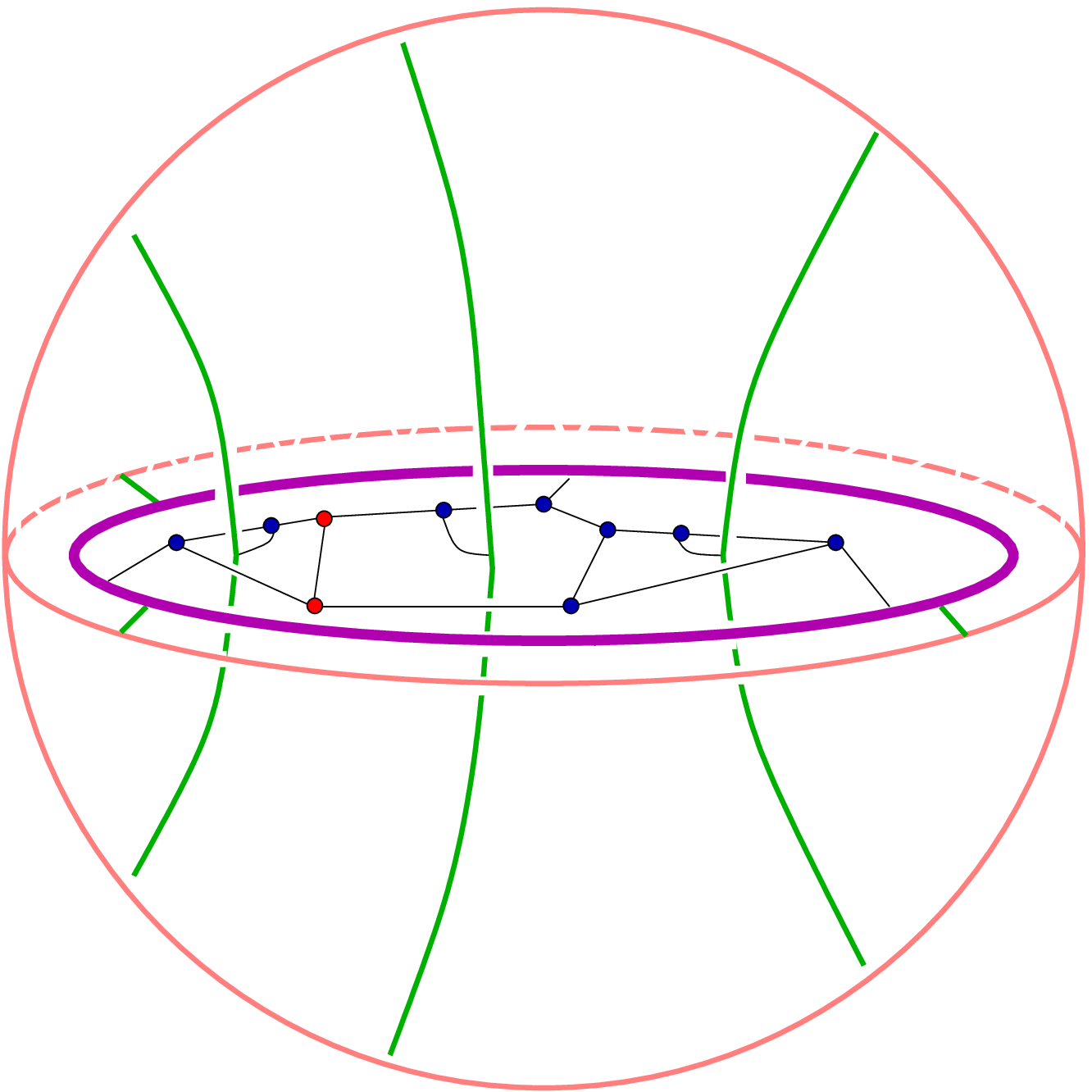}} \end{picture}}
  \put(77,53.7) {\scriptsize$M_1$}   \put(11.5,55) {\scriptsize$k_1$}
  \put(111,82)  {\scriptsize$M_2$}   \put(127,54)  {\scriptsize$k_2$}
  \put(-3.2,70) {\scriptsize$M_3$}   \put(12,85.6) {\scriptsize$k_3$}
  \put(20.7,29) {\scriptsize$j_1^+$} \put(22.6,112){\scriptsize$j_1^{-\vee}$}
  \put(55.4,8)  {\scriptsize$j_2^+$} \put(58.4,134){\scriptsize$j_2^{-\vee}$}
  \put(112.4,28){\scriptsize$j_3^+$} \put(111,111) {\scriptsize$j_3^{-\vee}$}
  \end{picture} $$ 
{}\\[-.96em]

\section{Results}

We have presented a complete and fully general prescription for all correlators 
on (orientable) world sheets of any topology. One can now prove that
this ansatz fulfils two types of consistency conditions: modular invariance
and factorization. First, the group {\sl Aut}$(\hat X,\sigma)$ of arc preserving
homeomorphisms of $\hat X$ of degree 1 that commute with the action of
$\sigma$ -- the `relative modular group' \cite2{bisa2} -- acts genuinely 
on the space $\calh(\hat X)$ of conformal blocks. The correlators on $\hat X$ 
can be shown to be invariant under this action. This general result contains
modular invariance of the torus partition function as a special case.
Second, we have compatibility with factorization: By gluing together two 
insertion points of a world sheet $X$ (either both of them in the bulk or both 
on the boundary) one gets
a world sheet $X'$ of different topology. We find that the 
correlators on $X$ and on $X'$ are correctly related by gluing maps.

\medskip

As special cases, our prescription for the correlators allows to compute 
explicitly partition functions and operator products. Besides being modular 
invariant, the partition function obtained for the torus without 
insertions can be shown to possess the usual integrality properties.
Similarly, the annulus without insertions yields a 
representation of the fusion rules by matrices with non-negative integral 
entries, a so-called NIM-rep. The one-point functions for bulk fields on the
disc give the boundary states. Their coefficients can be shown to furnish a 
classifying algebra \cite5{fuSc5}. In the simple current case one recovers
the boundary states proposed in \cite8{fhssw}. 
Finally, bulk and boundary OPEs can be computed.

For modular invariants of extension type, the category $\calc_{\!A}$ of 
$A$-re\-pre\-sen\-ta\-ti\-ons has additional structure. 
In particular $\calc_{\!A}$ is itself a tensor category, and it possesses
a `good' duality. In this situation, the annulus coefficients coincide with the 
fusion rules of $\calc_{\!A}$, the $6j$-symbols of $\calc_{\!A}$ give the 
boundary OPE, and the quantum dimension of an object in $\calc_A$ provides 
the boundary entropy \cite1{aflu} of the corresponding boundary condition.

\medskip

The algebra object $A$ is itself an $A$-representation and hence
corresponds to an (elementary) boundary condition. The algebra of open string 
states for this boundary condition (in string speak: the quantized algebra 
of functions on the corresponding brane) gives rise to the category theoretic 
algebra $A\,{\otimes_{\!A}}\,A\,{\cong}\,A$. We can therefore conclude:
\\[-1.1em]{}\begin{center} \fbox{\begin{minipage}{11.7cm}
The complete CFT can be reconstructed from the underlying chiral CFT by
using only the knowledge of the 
algebra of open string states for a {\em single\/} boundary condition.
\end{minipage}} \end{center}

This immediately raises the question whether such a reconstruction is
also possible from the algebra of open string states of {\em any}
elementary boundary condition. This is indeed the case: Every elementary 
boundary condition gives rise to an algebra object from which 
the {\em same\/} full CFT can be constructed. Therefore the algebra 
$A$ should {\em not\/} be thought of as an {\em observable\/} quantity -- 
a fact already familiar from lattice \tfts\ in two dimensions \cite{10}{fuhk}.
The algebras $A_1$ and $A_2$ corresponding to different boundary conditions 
are, however, closely related: they are Morita equivalent. This means that
there exist bimodules $_{A_1\!}M_{\!A_2}$ and $_{A_2\!}\tilde M_{\!A_1}$
(the first a left module of the algebra $A_1$ and a right module of $A_2$, 
and the second a left module of $A_2$ and a right module of $A_1$) such that
  $$  _{A_1\!}M_{\!A_2}\otimes_{\!A_2}^{} (_{A_2\!}\tilde M_{\!A_1}) = A_1
  \quad\ \mbox{and}\quad\
  (_{A_2\!}\tilde M_{\!A_1}) \otimes_{\!A_1}^{} (_{A_1\!}M_{\!A_2}) = A_2 \,.  $$
Together with orbifold techniques, Morita equivalence also allows for a
deeper understanding of T-duality.

\section{Conclusions}

Full rational conformal field theories that are based on the 
chiral data encoded in a modular tensor category $\calc$ can be
obtained from (Morita equivalence classes of) symmetric
special Frobenius algebras $A$ in $\calc$. We gave a general prescription
for the construction of correlation functions on orientable surfaces, including
surfaces with boundary. Boundary conditions are in one-to-one correspondence
to representations of $A$.

These results allow us to formulate two well-defined classification problems: 
the one of classifying all full CFTs based on a given chiral CFT, 
and the one of classifying all boundary conditions that preserve a certain 
subsymmetry of a given CFT, in the extreme case only conformal symmetry.
Except for the Virasoro minimal models, the latter situation is non-rational.
We are confident that the structures presented above are flexible enough to
be generalized to the non-rational case. In this regard,
we expect an intimate relation between the problem of deforming
Frobenius algebras in tensor categories and the problem of deforming conformal 
field theories. Our results should also be extended to include the case of 
unorientable surfaces, which appear naturally in type I string theories.  

\begin{acknowledgments} C.S.\ thanks the organizers of the
NATO Advanced Research Workshop on Statistical Field Theories for a
stimulating meeting and for the invitation to present this work.
Some of the results have been obtained in collaboration with Giovanni Felder 
and J\"urg Fr\"ohlich. We would like to thank them for discussions and for a 
very pleasant collaboration.  \end{acknowledgments}

 \newcommand\J[5]   {{#1}$\;${\bf #2},$\;${#4}$\;$({#3})}
 \newcommand\Prep[2]{preprint {#1}}
 \newcommand\PRep[2]{{#1}}
 \newcommand\inBO[7]{in:\ {\em #1} ({#3}, {#4} {#5}), p.\ {#6}}
 \newcommand\wb{\,\linebreak[0]} \def\wB {$\,$\wb}
 \newcommand\Bi[1]    {\bibitem{#1}}
 \renewcommand\Bi[3]  {\bibitem{#2}{\mbox{\hspace{.49em}}[#1]~#3}}
 \newcommand\BI[3]    {\bibitem{#2}{[#1]~#3}}
 \def\jf    {J.\ Fuchs}
 \def\bams  {Bull.\wb Amer.\wb Math.\wb Soc.}
 \def\comp  {Com\-mun.\wb Math.\wb Phys.}
 \def\fiic  {Fields\wb Inst.\wb Commun.}
 \def\ijmp  {Int.\wb J.\wb Mod.\wb Phys.\ A}
 \def\jgap  {J.\wb Geom.\wB and\wB Phys.}
 \def\jopa  {J.\wb Phys.\ A}
 \def\nuci  {Nuovo\wB Cim.}
 \def\nupb  {Nucl.\wb Phys.\ B}
 \def\phlb  {Phys.\wb Lett.\ B}
 \def\phrl  {Phys.\wb Rev.\wb Lett.}
 \newcommand\fscp[2] {\inBO{Fields, Strings, and Critical Phenomena} {E.\
     Br\'ezin and J.\ Zinn-Justin, eds.} \NH{Amsterdam}{1989} {{#1}}{{#2}}}
 \def\Ca     {{Cambridge}}
 \def\CUP    {{Cambridge University Press}}
 \def\NH     {{North Holland}}
 \def\NY     {{New York}}
 \def\PL     {{Plenum Press}}

\begin{chapthebibliography}{1}
\bibliographystyle{apalike}
\Bi1{aflu}    {I.\ Affleck and A.W.W.\ Ludwig, \J\phrl{67}{1991}{161} {Universal
              noninteger ``ground-state degeneracy" in critical \q systems}}
\Bi2{bisa2}   {M.\ Bianchi and A.\ Sagnotti, \J\phlb{231}{1989}{389}
              {Open strings and the relative modular group}}
\Bi3{fffs2}   {G.\ Felder, J.\ Fr\"ohlich, \jf, and C.\ Schwei\-gert,
              \J\phrl{84}{2000}{1659} {\Con \bc s and three-\dim al \tft}}
\Bi4{fffs3}   {G.\ Felder, J.\ Fr\"ohlich, \jf, and C.\ Schwei\-gert,
              pre\-print ETH-TH/99-30\,\&\,\\
              \mbox{$ $}\hspace{.55em}%
              PAR-LPTHE 99-45, Compos.\ Math., in press}
\Bi5{fuSc5}   {\jf\ and C.\ Schweigert,
              \J\phlb{414}{1997}{251} {A classifying \alg\ for \bc s}}
\Bi6{fuSc14}  {\jf\ and C.\ Schweigert, \J\phlb{490}{2000}{163} {Solitonic
              sectors, $\alpha$-induction and \sym\ breaking boundaries}}
\Bi7{fuSc16}  {\jf\ and C.\ Schweigert, \PRep{math.CT/0106050, to appear in \fiic}
              {Category theory for \con \bc s}}
\Bi8{fhssw}   {\jf, L.R.\ Huiszoon, A.N.\ Schellekens, C.\ Schweigert, and J.\
              Walcher, \J 
              {Phys.\\ \mbox{$ $}\hspace{.55em}Lett.\ B}{495}{2000}{427}
              {Boundaries, crosscaps and simple currents}}
\Bi9{fuRs}    {\jf, I.\ Runkel and C.\ Schweigert,
              preprint\,\,PAR-LPTHE\,01-45,\,\,hep-th/0110133}
\BI{10}{fuhk} {M.\ Fukuma, S.\ Hosono, and H.\ Kawai, \J\comp{161}{1994}{157}
              {Lattice \tft\ in two dimensions}}
\BI{11}{hora5}{P.\ Ho\v rava, \J\jgap{21}{1996}1
              {\CS gauge theory on orbifolds: open strings from three \dim s}}
\BI{12}{kios} {A.N.\ Kirillov and V.\ Ostrik, \Prep{math.QA/0101219} {On q-analog
              of McKay correspondence and ADE \class\ of $sl^{(2)}$ \cfts}}
\BI{13}{TUra} {V.G.\ Turaev, {\em Quantum Invariants of Knots and
              $3$-Manifolds} ({de Gruyter {1994}})}
\end{chapthebibliography} 
\end{document}